\documentclass[twocolumn, journal]{IEEEtran}
\usepackage{multirow}
\usepackage{dashrule}
\usepackage{amsmath}
\usepackage[OT2,T1]{fontenc}
\DeclareSymbolFont{cyrletters}{OT2}{wncyr}{m}{n}
\DeclareMathSymbol{\shah}{\mathalpha}{cyrletters}{"58}
\usepackage{amsfonts}
\usepackage{amssymb}
\usepackage{algorithmic}
\usepackage{mathrsfs}
\usepackage{array}
\usepackage{graphicx}
\usepackage{enumerate}
\usepackage{url}
\usepackage{cite}
\usepackage{tikz}
\usepackage{caption}
\ifCLASSOPTIONcompsoc
  \usepackage[caption=false,font=normalsize,labelfont=sf,textfont=sf]{subfig}
\else
  \usepackage[caption=false,font=footnotesize]{subfig}
\fi

\usepackage{comment}

\usepackage{bigdelim}
\usepackage{arydshln}

\newcommand{\JJ}{\mathcal{J}}
\newcommand{\II}{\mathcal{I}}
\newcommand{\KK}{\mathcal{K}}

\newcommand{\F}{\mathcal{F}}

\newcommand{\Z}{\mathbb{Z}}

\newcommand{\A}{\mathscr{A}}

\newcommand{\D}{\mathcal{D}}

\newcommand{\ZZ}{\mathcal{Z}}

\newcommand{\m}{\mathfrak{m}}
\newcommand{\vm}{\bar{\mathfrak{m}}}
\newcommand{\MC}{\mathcal{L}}
\newcommand{\Prob}{\mathfrak{i}}

\newcommand{\Gcdset}{\mathscr{A}}

\newtheorem{lemma}{Lemma}
\newtheorem{theorem}{Theorem}

\newtheorem{proposition}{Proposition}

\newtheorem{conjecture}{Conjecture}
\DeclareMathAlphabet{\mathpzc}{OT1}{pzc}{m}{it}
\def\x{{\mathbf x}}
\def\L{{\cal L}}

\renewcommand{\vec}[1]{{#1}}

\newcommand{\vh}{h}
\newcommand{\vx}{x}

\begin{document}

\title{Convolution Idempotents with a given Zero-set}

\author{Aditya~Siripuram,~\IEEEmembership{Member,~IEEE,}
        and Brad~Osgood,~\IEEEmembership{Member,~IEEE,}
\thanks{Aditya~Siripuram is with the Department of Electrical Engineering, IIT Hyderabad}
\thanks{Brad~Osgood is with the Department of Electrical Engineering at Stanford University.}
\thanks{\textcopyright 2019 IEEE.  Personal use of this material is permitted.  Permission from IEEE must be obtained for all other uses, in any current or future media, including reprinting/republishing this material for advertising or promotional purposes, creating new collective works, for resale or redistribution to servers or lists, or reuse of any copyrighted component of this work in other works.}
}

%



\maketitle
\begin{abstract}
We investigate the structure of $N$-length discrete signals $\vh$ satisfying $\vh*\vh=\vh$ that vanish on a given set of indices. We motivate this problem from examples in sampling, Fuglede's conjecture, and orthogonal interpolation of bandlimited signals. When $N=p^M$ is a prime power, we characterize \emph{all}  such $h$ with a prescribed zero set in terms of base-$p$ expansions of nonzero indices in $\F^{-1}h$. 
\end{abstract}


%
\IEEEpeerreviewmaketitle

\section{Introduction} 
\label{sec:intro}
A mapping $\vh:\mathbb{Z}_N \longrightarrow \mathbb{C}^N$ is a (convolution) \emph{idempotent} if $\vh*\vh = \vh$. Here $\mathbb{Z}_N$ are  the integers modulo $N$ and $*$ is circular convolution. We also regard discrete signals as vectors in $\mathbb{C}^N$, indexed from $0$ to $N-1$. This work deals with recovering $\vh$ when some of its elements are known to be zero. Our motivation for considering this comes from applications to several, apparently distant areas: multicoset sampling of analog signals (\cite{lin1998periodically}, Section  \ref{sec:motivation-multicoset}), Fuglede's conjecture  on spectral and tiling sets of integers \cite{fuglede1974commuting, siripuram2018lp, siripuram2019discrete}, and finding unitary submatrices of the discrete Fourier transform matrix \cite{siripuram2019discrete}.

Let us state:

\begin{quote}
    \emph{Zero-set Problem}: Given a positive integer $N$ and a set  $\ZZ \subseteq \mathbb{Z}_N$, find \emph{all} idempotents $\vh:\mathbb{Z}_N \longrightarrow \mathbb{C}^N$ that vanish on $\ZZ$. 
\end{quote}
 We let 
\[
\ZZ(\vh) = \{ n\in \mathbb{Z}_N : \vh(n)=0 \},
 \] 
and refer to it as the zero-set of $h$. For the zero-set problem we allow $\ZZ \subseteq \ZZ(h)$, \emph{i.e.},  $h$ has at least the zeros specified by $\ZZ$ but it may have more. In fact, a zero-set  $\ZZ(h)$ has an algebraic structure and cannot be arbitrary, and specifying that $h$ vanishes  on $\ZZ$ may force $h$ to have additional zeros, possibly up to $\Z_N$. In Section \ref{sec:zero-sets-of-idempotents} we will give a sharper formulation of the problem.  

In Section \ref{sec:idempotents-prime-power-case} we give a solution to the zero-set problem when the ambient dimension $N$ is a prime power. The proof uses basic Fourier analysis together with properties of Ramanujan sums. The latter have recently been utilized in signal processing, \cite{vaidyanathan2014ramanujan, vaidyanathan2014ramanujan2}.

We need a few notions and notations. We let $\F$ denote the discrete Fourier transform 
\[
\F \vx (n) = \sum_{k\in \Z_N}\vx(k)\omega_N^{-kn}, \quad 
\]
where $\omega_N = e^{2\pi/N}$ and $\vx:\mathbb{Z}_N \longrightarrow \mathbb{C}^N$.

An idempotent satisfies $(\F \vh (n))^2 = \F \vh (n)$, so a value $\F\vh(n)$ is either $0$ or $1$. Thus for a unique $\JJ \subseteq \Z_N$ we have
$\vh = \F^{-1} \underline{1}_\JJ$, \label{eq:h-j-defn}
where $\underline{1}_\JJ$ is the indicator function of $\JJ$. We write $h_\JJ$ 
when the dependence on $\JJ$ needs to be made explicit. 
Then 
$h_\JJ(n)=0$ when $\sum_{m\in \JJ} \omega_N^{mn}=0$,
and we see that an element of $\ZZ(h_\JJ)$ corresponds to a  vanishing sum of roots of unity. We also note  that if $\JJ = \JJ_1 \cup \JJ_2$ as a disjoint union then
$
h_\JJ = h_{\JJ_1}+h_{\JJ_2}.
$

We illustrate  solutions to the zero-set problem in the simple case $N=4$. 
 If $0 \in \ZZ (\vh) $ then $\sum_n \F \vh(n) = 0 $, and since $\F \vh(n) \in \{0,1\}$ for all $n$ this is only possible when $\vh = \vec{0}$. 
Of the seven remaining subsets of $\Z_N \setminus \{0\}$, we only need to consider $\ZZ (\vh)$ as one of $\{2\}$, $\{1,3\}$, and $\{1,2,3\}$; see Section \ref{sec:zero-sets-of-idempotents}. 

\begin{enumerate}
    \item When $\ZZ(\vh) = \{1,2,3 \}$, the idempotent $ \vh$ is a multiple of the discrete $\delta$, so clearly  $\F \vh = (1,1,1,1) = \mathbf{1}$, and $\vh = (1/4, 0, 0 ,0) = (1/4)\vec{\delta}$.
    \item When $\ZZ(\vh) = \{1,3\}$, $ \vh$ is obtained by upsampling a signal in $\mathbb{C}^2$ (\cite{osgood2018lectures}, \cite{oppenheim1999discrete}), and so the second half of $\F \vh$ must be a replica of the first half:\\
   $
    \F\vh \in \{(0,0,0,0), (1,0,1,0), (0,1,0,1), (1,1,1,1) \}
   $.
\item When \(\ZZ(\vh)=\{2\}\), we must have \(\F \vh (0)+ \F \vh (2) = \)\\ \(\F \vh (1) + \F \vh(3)\), and so
   \(    \F \vh \in \{(0,0,0,0), (1,0,0,1), \) \\ \((1,1,0,0), (0,0,1,1),(0,1,1,0), (1,1,1,1)\}\).

\end{enumerate}

\subsubsection{Bracelets}   On  $\Z_N$ we allow for operations of translation by $k$, $\tau^k(i)= i-k$, and reversal $\rho(i)= -i$.  Applying these operations, in any combinations, to an index set $\JJ\subseteq \mathbb{Z}_N$ yields the \emph{bracelet} of $\JJ$. (See also \cite{DS-1}.) For zero-sets:
\begin{proposition} \label{proposition:bracelet} 
If $\JJ$ and $\KK$ are in the same bracelet then $\ZZ(h_\JJ) = \ZZ(h_\KK)$.  
\end{proposition}
This follows at once from properties of the discrete Fourier transform on noting that $\underline{1}_{\tau^k\JJ} = \underline{1}_\JJ\circ \tau^{-k} $ and $\underline{1}_{\rho\JJ}=\underline{1}_\JJ \circ \rho$. Thus shifts and reversals of $\JJ$ do not change the zero-set $\ZZ(h_\JJ)$, and so the set of solutions to the zero-set problem must be closed for these operations; one sees this in the example above. This proposition is  useful in some of our arguments, for example allowing for a translation to assume that $0 \in \JJ$.

The converse to Proposition \ref{proposition:bracelet} is not true. For example, let $N=8$, $\JJ = \{0,1\}$, and $\KK = \{0,3\}$. One can check that $\ZZ(h_\JJ) = \{4\} = \ZZ(h_\KK)$, but $\JJ$ and $\KK$ are not in the same bracelet.

\section{Motivating Problems}
\label{sec:motivation}
We briefly discuss three very different scenarios that motivate the zero-set problem: multicoset sampling, Fuglede's conjecture, and unitary submatrices of the DFT. 
\subsection{Multicoset sampling}
\label{sec:motivation-multicoset}
  The traditional Nyquist-Shannon sampling theorem, using uniformly spaced samples, can be very inefficient for recovering signals with fragmented spectra. Nonuniform sampling techniques have been developed  that use frequency support information to develop a sampling pattern, see \cite{kohlenberg1953exact}, \cite{lin1998periodically} for some early work. The idea is to combine multiple uniform sampling patterns after appropriately shifting each of them (hence the term ``multicoset''). The sampling pattern for the discussion below is from \cite{kohlenberg1953exact}, \cite{lin1998periodically}, but tailored to the present discussion.
	
\begin{figure}[htb]
	\begin{center}
	\centering
	\begin{tikzpicture}[scale=1.2]
	\tikzstyle{every node}=[font=\small]
	\draw  (0,1.5) node [above left] {$\F f(s)$} -- (0,0);
	\draw  (-1,0) -- (3.5,0) node [below right] {$s$};
	\draw  (3,0) -- (3,1);
	\draw[domain=0:1] plot (\x, {1-\x^2});
	\draw[domain=2:3] plot (\x, {-8+6*\x-\x^2});
	\node [below ] at (0,0)  {$0$};
	\node [below ] at (1,0)  {$1$};
	\node [below ] at (3,0) {$3$};
	\node [below ] at (2,0) {$2$};
	\end{tikzpicture}
	\caption{Example signal with two fragments, for $\mathfrak{F} = \{0,2\}$.}
	\label{fig:nonunif-sampling-2}
	
	
	\end{center}
\end{figure}
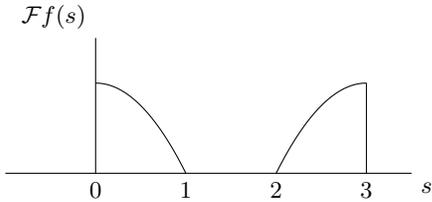

	For this section, we let $\F f$ denote the continuous-time Fourier transform of $f$. Consider the signal space $\mathscr{S}$ in which each signal has a \emph{fragmented spectrum}: \emph{i.e.} for any signal $f \in \mathscr{S}$, the Fourier transform $\F f (s)$ is nonzero only when $s\in \bigcup_{n \in \mathfrak{F}}[n, n+1] $. Here we assume the set $\mathfrak{F}$ consists  only of non negative integers. Each signal in $\mathscr{S}$ thus has a spectrum  consisting of $|\mathfrak{F}|$ fragments. For $k \in \mathfrak{F}$, we speak of the $k^\textsf{th}$ fragment to mean the fragment from $k$ to $k+1$. See Figure \ref{fig:nonunif-sampling-2} for an example signal with $\mathfrak{F}=\{0,2\}$ (\emph{i.e.} $2$ fragments), with the $0^\textsf{th}$ fragment in $[0,1]$ and $2^\textsf{nd}$ fragment in $[2,3]$.

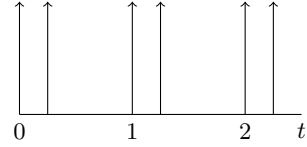
\begin{figure}[htb]
	\centering
	\begin{tikzpicture}[scale=1.5]
	\tikzstyle{every node}=[font=\small]
	\draw  (0,1.25) node [left] {};
	\draw  (0,0) -- (2.5,0) node [below] {$t$};
	\draw[->]  (0,0) node [below] {$0$} -- (0,1);
	\draw[->]  (0.25,0)  -- (0.25,1);
	\draw[->]  (1,0) node [below ] {$1$} -- (1,1);
	\draw[->]  (1.25,0) -- (1.25,1);
	\draw[->]  (2,0) node [below ] {$2$} -- (2,1);
	\draw[->]  (2.25,0) -- (2.25,1);
	\end{tikzpicture}
	\caption{Example sampling pattern in the case $\mathfrak{F}=\{0,2\}$, with the choice of $\JJ = \{0,1\}, N=4$}
	\label{fig:nonunif-sampling-1}
	
\end{figure}

Regular sampling of the signals in $\mathscr{S}$ at the Nyquist-Shannon sampling rate \cite{osgood2018lectures} might not make use of {gaps} in the fragmented spectrum, with the consequence that the sampling rate could be higher than required for reconstructing the signal. Consider the following construction of an irregular sampling pattern for signals in $\mathscr{S}$. Set $N > \max \mathfrak{F}  + 1$ (for example we could have $N=4$ for the spectrum in  Figure \ref{fig:nonunif-sampling-2}), and let 
\[
p_\JJ(t) = \sum_{m \in \JJ}\delta(t - m/N),
\]
where $\JJ \subseteq [0:N-1]$ is yet to be chosen. Consider sampling the signals in $\mathscr{S}$ with the sampling pattern 
\[
\sum_{k=-\infty}^\infty p_\JJ(t-kN) 
\]
and sampled signal 
\[
f_{\textsf{sampled}}(t) = f(t) \left(\sum_{k=-\infty}^\infty p_\JJ(t-kN)\right).
\]
See Figure \ref{fig:nonunif-sampling-1} for an example sampling pattern.

We find the spectrum of $f_{\textsf{sampled}}$ in the standard manner. Let $\shah(t) = \sum_n \delta(t-n)$, and note that the idempotent  $\vh_\JJ(n) = \F p_\JJ (n) $ is the $N$-point discrete Fourier transform  of $\underline{1}_\JJ$. Then  
\begin{align}
    \F f_{\textsf{sampled}}(s) &=  \F \left(f (\shah * p_\JJ) \right) (s) \nonumber \\
    &= \F f *\left(\shah\F \cdot p_\JJ  \right) (s)\nonumber \\
    &= \F f (s) * \left(\sum_{k=-\infty}^\infty \vh_\JJ(k)\delta(s-k) \right)\nonumber\\
    &= \sum_{k=-\infty}^{\infty} \vh_\JJ(k) \F f(s-k). \label{eq:multicoset-sampled-spectrum}
\end{align}
The following proposition links the viability of  the sampling scheme for recovering the signal to the zero-set of $h_\JJ$.
\begin{proposition}
\label{prop:multicoset-idempotent}
If $\vh_\JJ$ satisfies
$
\vh_\JJ(k_1-k_2) = 0$ whenever $k_1,k_2 \in \mathfrak{F}, k_1 \neq k_2$,
then $f$ can be recovered from the sampled spectrum in \eqref{eq:multicoset-sampled-spectrum}.
\end{proposition}
\begin{IEEEproof}
Write
\[
\F f_{\textsf{sampled}}(s) =  \vh_\JJ(0)\F f(s) + \underbrace{\sum_{k\neq 0} \vh_\JJ(k) \F f(s-k)}_\text{aliasing terms},
\]
and recall that $h_\JJ(0) \ne 0$. Now $\F f (s)$ is nonzero only when $s\in \bigcup_{n \in \mathfrak{F}}[n, n+1] $. 
For $k_1,k_2 \in \mathfrak{F}$, the  $k_2^\textsf{th}$ fragment collides with the $k_1^\textsf{th}$ fragment in the aliasing terms at a shift of $k=k_1-k_2$, but this contribution to the aliasing terms is scaled by $\vh_\JJ(k_1-k_2)$, which is $0$. For $k$ not of this form the shift $\F f(s-k)$ does not overlap with any fragment. Thus $\F f(s)$, hence $f$, can be recovered from the samples taken according to the sampling pattern $p_\JJ$.
\end{IEEEproof}

Observe that the average number of samples taken per second is $|\JJ|$, which could be much less than the Nyquist-Shannon rate. 

Given the signal space and the frequencies $\mathfrak{F}$, we wish to find an idempotent $\vh_\JJ$ that satisfies the conditions of Proposition \ref{prop:multicoset-idempotent}. Thus the problem of constructing a sampling pattern is related to the problem of finding an idempotent that vanishes on a given set.

\emph{Example}: 
Suppose $\mathfrak{F}= \{0,2\}$ as in Figure \ref{fig:nonunif-sampling-2}, and that a sampling pattern from Figure \ref{fig:nonunif-sampling-1} is used to sample the signal (this implies $\JJ = \{0,1\}, N = 4$). Note that 
\[
\vh_\JJ = \frac{1}{4}\begin{pmatrix}
2 & 1+i & 0 & 1-i
\end{pmatrix},
\]
and so $\vh_\JJ$ satisfies the hypothesis of Proposition \ref{prop:multicoset-idempotent}. Indeed, the spectrum of the sampled signal has the form
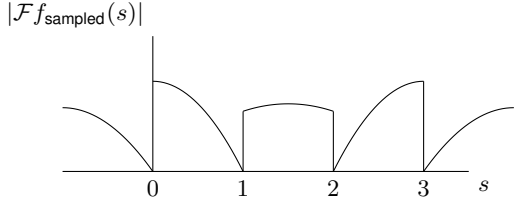
\begin{figure}[h]
	\begin{center}
	\centering
	\begin{tikzpicture}[scale=1.2]
	\tikzstyle{every node}=[font=\small]
	\draw  (0,1.5) node [above left] {$|\F f_\textsf{sampled}(s)|$} -- (0,0);
	\draw  (-1,0) -- (3.5,0) node [below right] {$s$};
	\draw  (3,0) -- (3,1);
	\draw[domain=0:1] plot (\x, {1-\x^2});
	\draw  (1,0) -- (1,0.6665);
	\draw[domain=1:2] plot (\x, {0.75 - 0.33*(\x-1.5)^2});
	\draw  (2,0) -- (2,0.6665);
	\draw[domain=2:3] plot (\x, {-8+6*\x-\x^2});
	\draw[domain=3:4] plot (\x, {(-8+6*(\x-1)-(\x-1)^2)/sqrt(2)});
		\draw[domain=-1:0] plot (\x, {(1-(\x+1)^2)/sqrt(2)});
	\node [below ] at (0,0)  {$0$};
	\node [below ] at (1,0)  {$1$};
	\node [below ] at (3,0) {$3$};
	\node [below ] at (2,0) {$2$};
	\end{tikzpicture}
\centering	\caption{Spectrum of signal in Fig \ref{fig:nonunif-sampling-2} sampled ($\JJ = \{0,1\}$, $\mathfrak{F} = \{0,2\}$)}
	\label{fig:nonunif-sampling-example}
	

	\end{center}
\end{figure}
The original fragments in the intervals $[0,1]$ and $[2,3]$ are unchanged, but the gaps between fragments are filled with aliases.

We have a few more comments.

\begin{enumerate}
    \item The constant $N$ is a design parameter of the technique. The imposed zero set is $\{k_1-k_2 | k_1\neq k_2 \in \mathfrak{F} \} \subseteq \Z_N$, and with a higher $N$ the relative size of the zero set is smaller, thus (hopefully) making it easier to find an idempotent that vanishes on this zero set. However, a large $N$ might also decrease the effective sampling rate (which is $|\JJ|$) and might also reduce the time interval between successive samples. 
    \item Here, for convenience,  we assumed that the width of all the spectral fragments is $1$. This can easily  be relaxed. 
    \item The technique uses the knowledge of the spectrum (via $\mathfrak{F}$) to construct the sampling pattern. Follow up works in  \cite{herley1999minimum} \cite{venkataramani2000perfect} try to remove this limitation and construct a \emph{universal} sampling pattern without any knowledge of the exact locations of the fragments. See \cite{DS-1} for a detailed treatment of universal sets. These techniques still require knowledge of the spectrum while reconstruction, and work in \cite{Mishali-Eldar} uses a compressed sensing (\cite{candes:robust}, \cite{vidyasagar2016tutorial}) based approach to make the reconstruction spectrum-blind.
\end{enumerate}


\subsection{Tiling and Fuglede's conjecture}
\label{sec:motivation-Fuglede}

Next, consider tiling sets in $\mathbb{Z}_N$. A set $\JJ \subseteq \mathbb{Z}_N$ \emph{tiles} $\mathbb{Z}_N$ if every $i\in \Z_N$ can be written uniquely as $i=j+k$ $\mod N$, with $j \in \JJ$ and  $k$ in a set $\KK \subseteq \Z_N$. More picturesquely, $\JJ$ together with its translates $\JJ+k$, $k \in \KK$ form a disjoint cover of $\Z_N$. Of course $\JJ$ and $\KK$ enter symmetrically, and writing
\[
(1_{\JJ}*1_{\KK})(n) = \sum_{j+k=n}1_{\JJ}(j)1_{\KK}(k)
\]
we see that the (symmetric) tiling condition is that $\JJ$ and $\KK$ satisfy
\begin{equation}
{1}_\JJ * {1}_\KK = {1_{\Z_N}} \label{eq:tiling-def}
\end{equation}

A set $\JJ \subseteq \Z_N$ is called a \emph{spectral set} if there exists a square unitary submatrix of $\F$ with columns indexed by $\JJ$. N.B., here and elsewhere in this paper when we say ``unitary'' we mean unitary up to scaling. 

 Fuglede's conjecture  for $\Z_N$, is:
\begin{conjecture}
	\label{conj:spectral<->tile}
	(Spectral iff Tiling) A set $\JJ \subseteq \Z_N$ is spectral if and only if $\JJ$ tiles $\Z_N$.
\end{conjecture}


%

In his original paper,  \cite{fuglede1974commuting},  Fuglede asks for the validity of this in a more general setting. A spectral set in Fuglede's sense is a domain $\Omega \subset \mathbb{R}^N$  for which there exists a \emph{spectrum} $\{\lambda_k\}_{k \in \mathbb{Z}} \subset \mathbb{R}^N$, meaning  that $\{e^{2\pi i \lambda_kx}\}_{k \in \mathbb{Z}} $ is an orthogonal basis for $L^2(\Omega)$. Then: 
\begin{conjecture}
	(Fuglede, \cite{fuglede1974commuting}): A domain $\Omega \subset \mathbb{R}^N$ is a spectral set if and only if it tiles $\mathbb{R}^N$.
	\end{conjecture}

Fuglede proved this to be true under the assumption that $\Omega$ is a lattice in $\mathbb{R}^N$. The conjecture has been disproved in the spectral $\implies$ tiling direction in $\mathbb{R}^3,\mathbb{R}^4,$ and  $\mathbb{R}^5$ \cite{tao2003fuglede, matolcsi2005fuglede, malikiosis2016fuglede}. The conjecture has also been disproved in the tiling $\implies$ spectral direction for $\mathbb{R}^3$ \cite{farkas2006fuglede}. The conjecture has also been proved to be true under more restrictive assumptions on the domain $\Omega$; for e.g. for convex planar sets \cite{iosevich2003fuglede} and union of intervals \cite{laba2001fuglede}. For cyclic groups $\Z_N$, such as we consider here,  the conjecture is known to be true in the case when $N$ is a prime power, see for example \cite{laba2002spectral}, \cite{taofugledeblog} and \cite{coven1999tiling}, and it has been proved in the Tiling $\implies$ Spectral direction when $N$ has at most two prime factors \cite{laba2002spectral, coven1999tiling}.  Other domains where Fuglede's conjecture is known to be true include the field of $p-$adic numbers \cite{fan2016compact, fan2015fuglede} and $\mathbb{Z}_p \times \mathbb{Z}_p$ \cite{iosevich2017fuglede}. See \cite{dutkay2014some} for the relationship between validity of conjectures in various domains. 

Consider the following direct approach to Conjecture \ref{conj:spectral<->tile}. Starting with a spectral set $\JJ$, to prove that $\JJ$ is a tiling set we need to find a $\KK$ such that \eqref{eq:tiling-def} holds. Then
\[
\vh_\JJ \vh_\KK = {\delta},
\]
and we need to find a $\KK$ such that $\vh_\KK$ vanishes on $\Z_N\setminus \{0\}$ wherever $\vh_\JJ$ does not:
\begin{equation}
\vh_\KK(n) = 0 \quad\text{for any }n\neq 0 \text{ such that }\vh_\JJ(n) \neq 0.
\label{eq:kk-tiling}
\end{equation}
Thus \eqref{eq:kk-tiling} asks us to find an idempotent $\vh_\KK$ that vanishes on a given set, and finding such an idempotent -- or the inability to find one -- would give insights into the validity of Conjecture \ref{conj:spectral<->tile}, at least in the direction spectral $\implies$ tiling.

\subsection{Unitary submatrices of the Fourier matrix}
A third problem is that of finding unitary  submatrices of the Fourier matrix $\F$: Find all possible rows $\II \subseteq \Z_N$ and columns $\JJ \subseteq \Z_N$ with $|\II| = |\JJ|$ such that the corresponding Fourier submatrix $M$ satisfies $M^*M = k I$, with $k$ a scalar. This is related to the problem of finding all orthogonal interpolating bases for  spaces of bandlimited signals, as treated in \cite{siripuram2019discrete}.

Here, since we are investigating the problem of enumerating all unitary submatrices, we will consider the \emph{enumeration complexity} \cite{johnson1988generating}, which is a measure of the delay between successive outputs, as a function of the input size. For the present problem (as earlier) we could define $\vh_\JJ = \F^{-1}\vec{1}_\JJ$. In our earlier work, we showed that it is possible to enumerate all possible $\II$, with a given zero set of $\vh_\JJ$, with constant enumeration complexity \cite{siripuram2019discrete}, by exploiting the structure of the graphs involved. The question then arises naturally as to the possibility of enumerating all $\JJ$, starting with a given zero set of $\vh_\JJ$. Finding all possible idempotents (and hence all possible $\JJ$) with a given zero set would have implications for constant time enumeration complexity of all unitary submatrices.

Additional challenges in formalizing the enumeration process include the order of enumeration and avoiding duplicates, which are not addressed in this paper.

\section{Algebraic Structure of Zero-sets, and Zero-set Divisors}
\label{sec:zero-sets-of-idempotents}

A foundational result is that the zero-set of an idempotent has an overall algebraic structure.   Let $\D_N$ be the set of all divisors of $N$ in $\Z_N$ (so omitting $N$), let $(i,N)$ denote  the greatest common divisor of $i$ and $N$, and let
	\begin{equation} \label{eq:A_N(k)}
	\A_N(k) = \{ i \in \Z_N \colon (i,N)=k\}.
	\end{equation}
Then

\begin{lemma} \label{lem:h-null}
	The zero-set $\ZZ(h)$ is the disjoint union 
	\[
	\ZZ(h) = \bigcup_{k\in \D(h)} \A_N(k)
	\]
	for the set of divisors $\D(h) =\ZZ(h) \cap \D_N$. 
\end{lemma}

We call $\D(h)$ the \emph{zero-set divisors of} $h$. It is also  helpful to describe $\ZZ(h)$ in terms of $\D(h)$ as
\begin{equation} \label{eq:zero-set-alternate}
\ZZ(h) = \{i \in \Z_N  \colon (i,N) \in \D(h)\}.
\end{equation}

The lemma appears in many different forms and contexts, see \cite{Will:thesis}, \cite{siripuram2019discrete}, or \cite[Theorem~2.1]{malikiosis2016fuglede} for example. We refer to these for the proof and background. 

 With  $\Z_{N/k}^\times$ denoting the multiplicative group of units in the ring $\Z_{N/k}$ (the elements in $\Z_N$ that are coprime to $k$) we have $\A_N(k)=k\Z_{N/k}^\times$, so
  $\ZZ(h)$ is essentially a disjoint union of multiplicative groups. It is because of Lemma \ref{lem:h-null} that we only needed to examine the possible zero-sets $\{2\}$, $\{1,3\}$, and $\{1,2,3\}$ in the example for $N=4$ in Section \ref{sec:intro}.

A converse of Lemma \ref{lem:h-null} would start with a disjoint union of multiplicative groups as above and ask to find an idempotent with that union as its zero-set. In fact, this cannot be done in all cases. For example, let $N=6$ and $\ZZ =\{2,3,4\}$. The set $\ZZ$ can be presented in the form given in the lemma, namely $\ZZ=\{2,4\} \cup \{3\}$, but an exhaustive search shows that there is no idempotent $h \in \mathbb{C}^6$ with $\ZZ(h) = \ZZ$.  We would know a great deal more  if we knew a general converse. 


We can now formulate a sharper form of the zero-set problem, and we do so in full generality.
\begin{quote}
	\emph{Problem $\Prob_N(\D)$}:  Given a positive integer $N$ and a set of divisors $\D \subseteq \D_N$ let 
	\begin{equation} \label{equation:zero-divisors}
	\ZZ=\{i \in \Z_N \colon (i,N) \in \D\} 
	\end{equation}
	Find all index sets $\JJ$ such that the idempotent $h_\JJ=\F^{-1}1_\JJ$ vanishes on $\ZZ$. 
\end{quote}

Obviously, if $\D' \subseteq \D$ then a solution to problem $\Prob_N(\D)$ is also a solution to problem $\Prob_N(\D')$. But also note that if $h$ is zero at a point in some $\A_N(k)\cap \ZZ$ then it is zero on all of $\A_N(k)$. This is a source of  possible ``extra'' zeros in a solution to a given zero-set problem.



\section{Solution to Problem $\Prob_N(\D)$ for $N$ a  prime power}
\label{sec:idempotents-prime-power-case}
Our goal is  to \emph{characterize all} index sets  that solve $\Prob_N(\D)$ when $N$ is a prime power. Our method relies on a systematic use of the base $p$ expansions of elements of the index set, together with properties of  Ramanujan's sum from number theory.

\subsection{Digit-tables}
 \label{section:digit-tables}
Everywhere in this section we assume that $N=p^M$ with $p$ prime. We write an index set  $\II = \{{i}_0, {i}_1, \dots\} \subseteq \Z_N$ in terms of the base-$p$ digits of its elements,   arrayed in rows. Thus associated to $\II$ is a table with entries in $[0:p-1]$, with $|\II|$ distinct rows, and with $M$ columns numbered from $0$ to $M-1$ giving the powers of $p$.  (So the leftmost digit in a row is  in the $1$'s place.) The order of the rows is not specified.   Conversely, with such rows thought of as base-$p$ expansions of elements of $\Z_N$ the table in turn determines an index set $\II \subseteq \Z_N$. 

Next, borrowing a term from matrix theory, we define the \emph{pivot columns} to be the columns which contain the first nonzero entry in some difference of rows. Precisely, let
\[
\begin{split}
 \MC &= \{j: \text{For some pair of rows $r$ and $r'$}\\
 & \text{one has    $i_{rj} \neq i_{r'j}$ and  $i_{rk} =  i_{r'k}$  for all  $k < j$} \}.
\end{split}
\]

The array of base-$p$ digits together with identified pivot columns constitute a \emph{digit-table}. 

We see that the initial columns of a digit-table, those prior to the first  pivot column, must each be constant, though the columns may be different constants. We also see that if there is a \emph{single} pivot column then \emph{all} the row differences must have first nonzero entry in that column, thus the entries in that column are distinct.

We use the notation $\m(M,\MC)$ for  a generic digit-table with $M$ columns  and with pivot columns indexed by $\MC$. We write $\m(\II, M,\MC)$ if we want also to identify the index set associated with the digit-table.

A first observation  is that the set of pivot columns is the same across the bracelet of an index set:
\begin{lemma} \label{lemma:pivot-bracelet}
    If  $\m(\II, M,\MC)$ and $\m(\II',M,\MC')$ are the digit-tables for two index sets in the same bracelet then $\MC=\MC'$. 
\end{lemma}
We omit the proof.

\subsection{Conforming Digit-Tables}
 \label{section:conforming-digit-tables}
In the definition of a digit-table we don't impose a condition on the number of rows, \emph{i.e.}, on the size of the associated index set. But a crucial such  condition arises for the digit-tables that give solutions of the zero-set problem. We say that $\m(M,\MC)$ is a \emph{conforming digit-table} if the number of rows is $p^{|\MC|}$. To indicate a conforming digit-table we use the notation $\vm(M,\MC)$, or $\vm(\II,M,\MC)$, with a bar.

 Conformity has an important consequence for the structure of a digit-table, and in turn for the structure of solutions to the zero-set problem.
 
 \begin{lemma} \label{lemma:conforming-structure-lemma}
 Let $\vm(\II,M,\MC)$ be a conforming digit-table, with $\MC = \{l_0, l_1, \dots,l_k \}$ and the entries of $\MC$ labeled in ascending order. Let $\MC_1=\MC \setminus \{l_0\}$. Arrange the rows of $\vm(\II,M,\MC)$ lexicographically. Then $\vm(\II,M,\MC)$ is a concatenation of $p$ disjoint, conforming digit-tables $\vm(\JJ_0,M,\MC_1), \dots, \vm(\JJ_{p-1},M,\MC_1)$, as illustrated in Table \ref{tab:hnull-converse-block-ordering-1}. 
     \end{lemma}

\begin{table}[h]
	\begin{center}
		\begin{tabular}{ccc:c:ccc}
			
			$p^0$ & $p^1$ & $\cdots$ & $p^{l_0}$ & $p^{l_0+1}$ & $p^{l_0 +2}$ \, $\cdots$ \, $p^{M-1}$\\
			\cline{1-6}
			\hspace{-.45in}\ldelim\{{13.5}{*}[$a$] & & & 0 & & &  \hspace{-.15in}\rdelim\}{4}{*}\\ 
			& & & \vdots & & \hspace{-.24in}$\vm(\JJ_0,M,\MC_1)$  & \\
			& & & 0 & & & \\
			\cdashline{1-6}
			& & & $1$ & & & \hspace{-.15in}\rdelim\}{4}{*} \\   
		    & & & \vdots & & \hspace{-.24in}$\vm(\JJ_1,M,\MC_1)$ &  \\
			& & & $1$ & &  & \\
			\cdashline{1-6}
			& & & \vdots & & & \\
	        \cdashline{1-6}
	        & & & $p-1$ & & & \hspace{-.15in}\rdelim\}{4}{*}\\
		    & & & \vdots & &  $\vm(\JJ_{p-1},M,\MC_1)$ & \\
			& & & $p-1$ & & & 
		\end{tabular}
		\vspace{2pt}
		\caption{The structure of a conforming digit-table $\vm(\II,M, \MC)$}
		
		\label{tab:hnull-converse-block-ordering-1}
	\end{center}
\end{table}

In the table,  $a$ is a row vector of the digits in the first $l_0-1$ columns in the lexicographic ordering  of the rows of $\vm(\II,M,\MC)$. Recall that each of these columns is constant, so $a$ is repeated down the rows of the table.

The issue is the conformity of the $p$ smaller digit-tables. The sizes of the smaller digit-tables must sum to $p^{|\MC|}$ and we claim that each must  have $p^{|\MC|-1}$ rows.

\begin{IEEEproof}  
For any row in any of the smaller digit-tables, there are at most $p^{|\MC|-1}$ distinct values for the  digits in the columns $\MC_1$. Suppose one of the smaller digit-tables has more than $p^{|\MC|-1}$ rows. By the pigeon hole principle, at least two of these rows must have the same entries in all the columns in $\MC_1$, and also, since the rows are in the same smaller digit-table, their entry in the column $l_0$ is also the same. This means that when we take the difference of these two rows the columns in $\MC$ all give a difference of $0$, whence the pivot column is outside $\MC$. That is a contradiction.
\end{IEEEproof}

Note that for a digit-table with a single pivot column $l_0$ the smaller conforming digit-tables are single rows,  and the $l_0$-column is just the digits from $0$ to $p-1$.

\subsection{Structure of idempotents with a given zero-set}
The following result identifies a general solution to the zero-set problem for a prime power $N=p^M$. To emphasize the connection to digit-tables we write the zero-set divisors from \eqref{equation:zero-divisors} as
\[
\D = \{p^{l_0}, p^{l_1}, \dots, p^{l_{k-1}}\} =: p^\MC, 
\]
where $\MC=\{l_0,\dots, l_{k-1}\} \subseteq [0:M-1]$. We also introduce a set of powers derived from $\MC$,
\begin{equation} \label{eq:MC^*}
 \MC^* = M - \MC - 1.
\end{equation}
 
\begin{theorem} \label{theorem:general-converse-hnull}
 An index set $\JJ$ is a solution to $\Prob_N(p^\MC)$ if and only if the digit-table for $\JJ$ is a concatenation of disjoint, conforming digit-tables:
	\begin{equation*}
	\begin{pmatrix}
	\vm(\II_0,M, \MC^*) \\ 
	\vm(\II_1, M,\MC^*) \\ 
	\vdots
	\end{pmatrix}.
	\end{equation*}
  
\end{theorem}

The index sets $\II_\nu$, all of cardinality $p^{|\MC^*|}=p^{|\MC|}$, are disjoint, with $\JJ =\II_0 \cup \II_1 \cup \cdots$. (In particular $|\JJ|$ is divisible by $p^{|\MC|}$.) The idempotent $h_\JJ$ is the sum
$
h_\JJ= \sum_\nu h_{\II_\nu}
$.


The proof of both necessity and sufficiency in Theorem \ref{theorem:general-converse-hnull}  will be by induction on the cardinality $|\MC|$. 
We begin with sufficiency. 

We first establish that an index set associated with a single conforming digit-table $\vm(\II_0,M,\MC^*)$ is a solution to $\Prob(p^{\MC})$. Suppose $\MC^*= \{l^*\}$ is a singleton -- the first step in the induction. This is 
the case of a single pivot column, and  Lemma \ref{lemma:conforming-structure-lemma} for $\vm(\II_0,M, \MC^*)$ takes the form
\begin{table}[h]
	\begin{center}
		\begin{tabular}{ccc|c|ccc}
			
			$p^0$ & $p^1$ & $\ldots$ & $p^{l^*}$ & $p^{l^*+1}$ & $\ldots$ & $p^{M-1}$\\
			\cline{1-7}
			\hspace{-.4in}\ldelim\{{6}{*}[$a$] & & & 0 &  & $b_0$  \\
			\multicolumn{3}{c|}{} & $1$ &   & $b_1$  \\ 
			\multicolumn{3}{c|}{} & $2$ &   & $b_2$ \\
			\multicolumn{3}{c|}{} & $\vdots$ &  & $\vdots$  \\
			& & & $p-1$ &  & $b_{p-1}$  \\
			\hline
			\end{tabular}.
	\end{center}
\end{table}

\noindent More explicitly, the elements in $\vm(\II_0,M, \MC^*)$ can be written 
\[
a + \{0\cdot p^{l^*} + b_0, 1\cdot p^{l^*} + b_1, 2\cdot p^{l^*} + b_2, \ldots, (p-1)p^{l^*} + b_{p-1} \}, 
\]
where $b_0, b_1, \ldots, b_{p-1}$ are multiples of $p^{l^*+1}$.


The corresponding idempotent is
\[
\begin{aligned}
h(n) &= \frac{1}{N}\sum_{j=0}^{p-1}e^{\left(2\pi i n(a + j p^{l^*} + b_j)/N\right)}\\
&= \frac{e^{2\pi ni a/p^M}}{p^M}\left(\sum_{j=0}^{p-1} e^{2\pi i (nj p^{l^*} + nb_j)/p^M}\right).
\end{aligned}
\]
Evaluating $h$ at $n = \alpha p^{M-l^*-1}$, where $\alpha$ is coprime to $p$, we get
\[
h(\alpha p^{M-l^*-1}) = \frac{e^{2\pi \alpha i a/p^{l^*+1}}}{p^M}\left(\sum_{j=0}^{p-1} e^{2\alpha \pi i j /p}\right) = 0.
\]
Thus $\II_0$ provides a solution to $\Prob(p^{M-\{l^*\}-1})=\Prob(p^{\MC})$. 

Now consider a conforming digit table $\vm(\II_0,M, \MC^*)$, $\MC^* = \{l_0^*,l_1^*,\dots, l_{k-1}^*\}$. As in Lemma \ref{lemma:conforming-structure-lemma} label the elements of $\MC^*$ is ascending order, and with $\{l_0,l_1, \dots, l_{k-1}\} = \MC = M-\MC^*-1$ labeled in descending order. 
Let $\MC_1^* = \MC^* \setminus \{l_0^*\}$.  By Lemma \ref{lemma:conforming-structure-lemma} applied to $\MC^*$, the digit-table $\vm(\II_0,M, \MC^*)$ splits into smaller digit-tables as in Table \ref{tab:hnull-converse-block-ordering-1}, and  the set $\II_0$  splits into the union of $p$ sets, 
\begin{equation} \label{eq:I_0-induction}
\begin{aligned}
    \II_0 &=a +  \left(0\cdot p^{l_0^*} + \JJ_0\right) \cup\left(1 \cdot p^{l_0^*} + \JJ_1\right)\cup\\
    & \cdots \cup \left((p-1)\cdot p^{l_0^*} + \JJ_{p-1}\right),
    \end{aligned}
    \end{equation}
where:

\begin{enumerate}
    \item Each $\JJ_k$ corresponds to a conforming digit-table with pivot columns $\MC_1^*$. By the induction hypothesis, each $\JJ_k$ is a solution to $\Prob(p^{\MC_1})$, $\MC_1=M-\MC_1^*-1$, or in other words $h_{\JJ_k}(n) = 0$ for $n \in \bigcup_{l \in \MC_1}\Gcdset(p^{l})$. 
    \item Each $\JJ_k$ contains only elements  that are multiples of $p^{l_1^*}$. Thus the idempotent $\vh_{\JJ_k}(n)$ when evaluated at multiples $\beta p^{M-l_1^*}$ of $p^{M-l_1^*}$ results in 
    \begin{equation}
    \begin{aligned}
    \label{eq:conv-hnull-converse-1}
    \vh_{\JJ_k}(\beta p^{M-l_1^*}) &= \frac{1}{N}\sum_{m \in \JJ_k}e^{2\pi i m \beta p^{M-l_1^*}/N}\\
    & = \frac{1}{N}\sum_{m \in \JJ_k} 1 = |\JJ_k|/N = p^{|\MC^*|-1}/N.
    \end{aligned}
    \end{equation}
In particular, since $p^{M-l_0^*}$ is a multiple of $p^{M- l_1^*}$,  we have that $\vh_{\JJ_k}(\beta p^{M-l_0^*}) = p^{|\MC^*|-1}/N$.
\end{enumerate}
Now consider the idempotent $\vh_{\II_0}$, where according to \eqref{eq:I_0-induction}:
\[
\begin{aligned}
\vh_{\II_0}(n) &= \frac{1}{N}\sum_{m\in \II_0}e^{2\pi i mn/N}\\
&= \frac{e^{2\pi i a n / N}}{N}\sum_{k=0}^{p-1}e^{2\pi i k n p^{l_0^*}/N}h_{\JJ_k}(n).
\end{aligned}
\]
As per 1) above, for $n \in \bigcup_{l \in \MC_1}\Gcdset(p^{l})$ the idempotent evaluates to $0$ by the induction hypothesis on the $h_{\JJ_k}$. While for $n \in \Gcdset(p^{l_0})$, say $n=\alpha p^{l_0} = \alpha p^{M-l_0^\star -1}$, with $\alpha$ coprime to $p$, from \eqref{eq:conv-hnull-converse-1} we get
\begin{equation}
\vh_{\II_0}(\alpha p^{l_0}) = \frac{p^{|\MC^*|-1}}{N}e^{2\pi i a \alpha p^{l_0} / N}\sum_{k=0}^{p-1}e^{2\pi i k \alpha/p} = 0.
\end{equation}
Thus $h_{\II_0}(n)$ vanishes at $n \in \bigcup_{l \in \MC}\Gcdset(p^{l})$, which proves that $\II_0$ solves $ \Prob(p^\MC)$. 

This completes the induction for a single digit-table $\vm(\II_0,M,\MC^*)$. A concatenation of such conforming digit tables yields a set of the form 
\(
\JJ = \II_0 \cup \II_1 \cup \II_2 \cup \ldots, 
\)
a disjoint union of sets that come from the individual conforming digit-tables. Then $ h_\JJ = \sum_k h_{\II_k}$,  and since each of the  $h_{\II_k}$ vanish on $\bigcup_{l \in \MC} \A_N(p^{M-l-1})$ so does $h_\JJ$, meaning that $\JJ$ is a solution to $\Prob(p^{\MC})$. This establishes sufficiency in Theorem \ref{theorem:general-converse-hnull}.

Next, necessity.  To prepare for the proof, first observe that a natural way to characterize an $h$ with $\ZZ$ as a zero-set is to specify that $h \cdot {1}_\ZZ  = 0$. So if $h$ is an idempotent with $h = \F^{-1}1_\JJ$   for an index set $\JJ$ and $c = \F^{-1} {1}_\ZZ$,
then $  1_{\JJ^-}* c = 0$ or 
\begin{equation}
\label{eq:ram-sum-0}
\sum_{j \in \JJ} c(n+j) = 0,
\end{equation}
for any $n \in \Z_N$. Here $\JJ^- = \{-j : j \in \JJ\}$, and we used $N\F^{-1}\F^{-1}1_\JJ(n) = 1_\JJ(-n)$.  
Finding solutions to the problem $\Prob_N(p^\MC)$ {is equivalent to finding index sets $\JJ$ satisfying \eqref{eq:ram-sum-0}}. What is the structure of such a set? That is the upshot of the necessity in Theorem \ref{theorem:general-converse-hnull}.

First suppose that $\D = \{p^l\}$, the initial step in the induction. The corresponding zero-set is a single one of the groups $\A_N(k)$, namely 
\begin{equation} \label{eq:putative-zeros}
\ZZ = \{i\in \Z_N : (i,p^M) = p^l \} = \A_{p^M}(p^l)= \bigcup_{(\alpha,p^M)=1}\{\alpha p^l\}.
\end{equation}
 It is through a connection to Ramanujan's sum, from number theory, that we are able to proceed.  Ramanujan's sum is 
\begin{equation}
\label{eq:def-ramanujan}
\mathfrak{c}_q(k) = \sum_{\substack{n\in \Z_q \\ (n, q) = 1}}\cos 
({2\pi nk}/{q})=\sum_{\substack{n\in \Z_q \\ (n, q) = 1}}\exp ( 2 \pi i n k / q ),
\end{equation}
where $q$ and $k$ are positive integers. See, for example, the original paper \cite{ramanujan1918certain}.   We will need the following two properties:

\begin{enumerate}[(i)]
	\item When $q=p^m$ is a prime power, 
	\begin{equation}
	\label{eq:ramanujan-sum-exp}
	\mathfrak{c}_{p^m}(k) = \begin{cases}
	0& \quad  \text{ if }p^{m-1} \nmid k, \\
	-p^{m-1}& \quad  \text{ if }p^{m-1} \ | \ k \text{ and }p^m \nmid k,\\
	\phi(p^m)& \quad \text{ if } p^m \ | \ k,
	\end{cases}
	\end{equation}
	where $\phi$ is the Euler totient function.	
	\item For any divisor $d$ of $q$,
	\begin{align*}
	&\sum_{\substack{n \in \Z_q \\ (n, q) = d}}\exp (2\pi i nk /q)  =\mathfrak{c}_{d'}(k).
	\end{align*}
	where $d' = q/d$.
\end{enumerate}

From \eqref{eq:putative-zeros} and \eqref{eq:def-ramanujan}, and from property (ii) with 
\[
q=N=p^M,\quad  d=p^l, \quad \text{and} \quad
d'= p^{M-l}=N/d,
\] 
we can identify $\mathfrak{c}_{d'}$ as an inverse Fourier transform
$
N\F^{-1} {1}_\ZZ = \mathfrak{c}_{d'}$. 

We can then rewrite  \eqref{eq:ram-sum-0} as the condition that
\begin{equation}
\label{eq:h-null-ft-condition}
\sum_{j \in \mathcal{K}} \mathfrak{c}_{d'} (j) = 0,  
\end{equation}
holds where $\mathcal{K}$ is any index set in the bracelet of $\JJ$. 
Forming the bracelet of $\JJ$ includes reversals, $\JJ^- = \{-j : j \in \JJ\}$, but we can regard $\mathfrak{c}_{d'}(k)$ to be defined for all integers $k$, and, from \eqref{eq:def-ramanujan}, it is even.   
We will use property (i) to deduce the structure of a digit-table of an index set $\JJ$ satisfying \eqref{eq:h-null-ft-condition}. Recall from Lemma \ref{lemma:pivot-bracelet} that  forming the bracelet does not change the pivot columns of the digit-table.  

Let 
$ l'=M-l$, 
so $d'=p^{l'}$, and we invoke property (i) with $m=l'$. Translate $\JJ$ to assume that $0 \in \JJ$.  
The values of $\mathfrak{c}_{d'}$ are integers (a general fact, but in our case it follows from property (i)),  and
$\mathfrak{c}_{d'}(0) = \phi(d') = (p-1)p^{l'  - 1}$.
Since $\mathfrak{c}_{d'}$ is periodic with period $d'$, we get
\[
(p-1)p^{l'-1} = \mathfrak{c}_{d'}(0) = \mathfrak{c}_{d'}(d') = \mathfrak{c}_{d'}(2d') =\ldots.
\]
These are the only positive values of $\mathfrak{c}_{d'}$.
Now, again from property (i), since a negative value of $\mathfrak{c}_{d'}$ can only be $-p^{l' -1 }$,  we need at least $p-1$ of these negative values to cancel out one positive value, and   each positive value occurring among the values in $\mathfrak{c}_{d'}$ necessitates the occurrence of $p-1$ negative values. 
Property (i) also says that the negative values of $\mathfrak{c}_{d'}$ are at indices that are multiples of $ p^{l'-1}$, excluding those that are multiples of $p^{l'}$. So $\JJ$ includes, say,  $r$ multiples of $p^{l'}$  (these are the indices where the value of $\mathfrak{c}_{d'}$ is positive) and $r(p-1)$ multiples of $p^{l'-1}$ that are not multiples of $p^{l'}$ (these are the indices where the value of $\mathfrak{c}_{d'}$ is negative). 

In summary, we have now determined that $\JJ$ has to include certain indices all of which are multiples of $p^{l'-1}$. Isolate these indices as a subset $\JJ_1$ of $\JJ$. Then the bracelet of $\JJ_1$ satisfies \eqref{eq:h-null-ft-condition}, for a translation of $\JJ_1$ either results in the same set (if the translation is by a multiple of $p^{l'-1}$), or results in a set all of whose indices are not divisible by $p^{l'-1}$. In either case \eqref{eq:h-null-ft-condition} holds. For reversal of $\JJ_1$  we rely on the fact that $\mathfrak{c}_{d'}$ is even.  

In addition to these indices, $\JJ$ could contain indices $k$ such that $p^{l'-1} \nmid k$, so that $\mathfrak{c}_{d'}(k) = 0$, and  \eqref{eq:h-null-ft-condition} is unaffected. In this last  case we can isolate all such $k$ into another subset $\JJ'$ of $\JJ$.  In this context we make the following simple, general observation:

\begin{itemize}
\item 	If $h = \F^{-1} 1_\JJ$ and $h_1 = \F^{-1} 1_{\JJ_1}$ are both solutions to $\Prob_N(\D)$, and if $\JJ_1 \subseteq \JJ$, then $h' = \F^{-1} 1_{\JJ \setminus \JJ_1}$ is also a solution to $\Prob_N(\D)$.
\end{itemize}

So we see that $\JJ'=\JJ\setminus\JJ_1$ itself gives a solution to \eqref{eq:h-null-ft-condition}. We can translate $\JJ'$ so that $0 \in \JJ'$, and repeat the arguments above with $\JJ'$ instead of $\JJ$. But then $\JJ'$ (appropriately translated) contains the indices described previously, and hence can be broken down further. Repeating this process, it follows that $\JJ$ breaks down into a disjoint union
$\JJ = \JJ_1 \cup \JJ_2 \cup \JJ_3 \cup \cdots$,
where each  $\JJ_i$, when appropriately translated, contains only multiples of $p^{l'-1}$. 

For the remainder of the argument we can thus assume that $\JJ$ contains only multiples of $p^{l'-1}$; arbitrary solutions can be constructed by taking disjoint unions of (translates of) such sets.

Under this assumption it would seem that a natural way to study $\JJ$ is to reduce it modulo $d'=p^{l'}$. Write $\JJ/d'$ for the set of congruence classes of elements of $\JJ$  mod $d'$ and write
$(\JJ/d')^{\sim}$ for the corresponding {multiset} where congruence classes are listed according to their multiplicity.   
  Then $(\JJ/d')^{\sim}$ has $r$ zeros and $r(p-1)$ non-zeros, where the non-zeros are multiples of $p^{l'-1}$.  (Intriguingly, this must hold for \emph{any translate} of $\JJ$.)  Using this, we will argue that $(\JJ/d')^{\sim}$ contains all non-zero elements in \emph{equal} amounts, specifically that  each nonzero multiple of $p^{l'-1}$ in $(\JJ/d')^\sim$  has multiplicity $r$.

Suppose the multiset is
\[
(\JJ/d')^\sim =  \begin{Bmatrix}
0 & \text{ with multiplicity } & r, \\
\alpha p^{l'-1} &  \text{ with multiplicity } & r_1, \\
\vdots & \vdots & \vdots
\end{Bmatrix},
\]
where $\alpha \in [1:p-1]$, and $\alpha p^{l'-1}$ is the non-zero element with highest multiplicity. Note that there are $p-1$ non-zero entries, and the multiplicities of the non-zero entries must add up to $r(p-1)$. Now suppose we translate $\JJ$ by  $\beta =-\alpha p^{l'-1}$, denoting the translated index set by $\tau^\beta\JJ$. Then the multiset $(\tau^\beta\JJ/d')^\sim$ contains $0$ with multiplicity $r_1$, so the multiplicity of remaining elements must add up to $r_1(p-1)$, which is impossible unless $r_1 = r$.

We summarize our analysis as follows: 

\begin{lemma}
	\label{lem:conv-singleton-case}
	For any solution $\JJ$ to to the singleton problem $\Prob_N(p^{l'})$ such that $0 \in \JJ$, we have
	$(\JJ/d')^\sim =p^{l'-1}\{0, \dots,1, \dots, 2,\dots, p-1, \dots \}$ 
	where all the multiplicities are equal (to $r$).
\end{lemma}

Interpreting this 
in terms of digit-tables, the digit-table for such a $\JJ$ is of the form

\begin{center} 
	\begin{tabular}{ccc|c|c}
		
		$p^0$ & $p^1$ & $\ldots$ & $p^{l'-1}$ & $\ldots$ \\
		\cline{1-5} 
		$0$ & $0$ & $\ldots$ & $0$ & $\ldots$ \rdelim\}{5.5}{2mm}[ $\vm(\II_0,M,\MC^*)$]\\
		$0$ & $0$ & $\ldots$ & $1$ & $\ldots$ \\ 
		$0$ & $0$ & $\ldots$ & $2$ & $\ldots$ \\
		$\vdots$ & $\vdots$ & $\vdots$ & $\vdots$ & \vdots \\
		$0$ & $0$ & $\ldots$ & $p-1$ & $\ldots$ \\
		\cline{1-5}
		$0$ & $0$ & $\ldots$ & $0$ & $\ldots$ 
		\rdelim\}{5.5}{2mm}[ $\vm(\II_1,M,\MC^*)$]\\
		$0$ & $0$ & $\ldots$ & $1$ & $\ldots$\\ 
		$0$ & $0$ & $\ldots$ & $2$ & $\ldots$ \\ 
		$\vdots$ & $\vdots$ & $\vdots$ & $\vdots$ & \vdots \\
		$0$ & $0$ & $\ldots$ & $p-1$ & $\ldots$\\
		\cline{1-5}	
		$0$ & $0$ & $\ldots$ & $0$ & $\ldots$ 
		\rdelim\}{5.5}{2mm}[ $\vm(\II_2,M,\MC^*)$]\\
		$0$ & $0$ & $\ldots$ & $1$ & $\ldots$\\ 
		$0$ & $0$ & $\ldots$ & $2$ & $\ldots$ \\  
		$\vdots$ & $\vdots$ & $\vdots$ & $\vdots$  & \vdots \\
		$0$ & $0$ & $\ldots$ & $p-1$ & $\ldots$\\
		\cline{1-5}
		$\vdots$ & $\vdots$ & $\vdots$ & $\vdots$ &
	\end{tabular}
	
\end{center}

\noindent where the digits in the $(l' -1)^{th}$ column contains all the digits $\{0,1,2\ldots, p-1\}$ in equal amounts, and the digits from column $l'$ onward may be arbitrary, other than assuring that the rows are distinct. 

Stated differently, each of the $r$ blocks is a conforming digit-table with $p=p^{|\MC^*|}$ rows, with $\MC^*=M-1-\MC=\{l'-1\}$, and with a single pivot column $l' - 1$.  That is, the digit-table for $\JJ$ is of the type
\begin{equation*}
\begin{pmatrix}
\vm(\II_0,M, \MC^*) \\ 
\vm(\II_1, M, \MC^*) \\ 
\vdots
\end{pmatrix}.
\end{equation*}
This is precisely the form that Theorem \ref{theorem:general-converse-hnull} takes when $\MC$ is a singleton, and thus establishes the first step in the induction.

As a side comment, it is possible to translate $\JJ$ by any amount, so the first $l'-2$ columns need not be all zeros -- this does not affect the pivot column. It is also possible to take a disjoint union with other sets of this type; note that disjoint union of sets corresponds to simply concatenating the digit-tables row-wise.

For the induction step, let $\MC = \{l_0, l_1, \ldots, l_{k -1}\}$, where the entries are labeled in ascending order. 
We examine a digit-table for any solution $\JJ$ to $\Prob_N(p^\MC)$. Any solution to $\Prob_N(p^\MC)$ is also a solution to $\Prob_N(p^{\mathcal{K}} )$ for any subset $\mathcal{K}$ of $\MC$, thus, in particular, the digit-table gives a solution to the problem $\Prob_N(p^{\{l_{k-1}\}})$. Suppose the rows of the digit-table for $\JJ$ are arranged lexicographically. By the result for a singleton,  the first pivot column of the digit-table is $l_0^\star = M - l_{k-1} - 1$ and the digit-table splits into blocks depending on the values  up to the $l_0^\star$ column, as in Table \ref{tab:conv-hnull-generic-ordering}. In each block the $a$'s are row vectors of the digits in the first $l_0^*-1$ columns.
\begin{table}[h]
	\begin{center}
		\begin{tabular}{ccc|c|cc}
			
			$p^0$ & $\cdots$ & $\cdots$ & $p^{l_0^*}$ & $p^{l_0^*+1}$ & $\cdots$\\
			\cline{1-6}
			\multicolumn{3}{c|}{${a}_0$} & $\ldots$ & $\ldots$ & $\ldots$\\
			\multicolumn{3}{c|}{${a}_0$} & $\ldots$ & $\ldots$ & $\ldots$\\ 
			& $\vdots$ & & $\ldots$ & $\ldots $ & $\ldots$\\
			\hline
			
			\multicolumn{3}{c|}{${a}_1$} & $\ldots$ & $\ldots$ & $\ldots$\\
			\multicolumn{3}{c|}{${a}_1$} & $\ldots$ & $\ldots$ & $\ldots$\\
			& $\vdots$ &  & $\ldots$ &  $\ldots$ & $\ldots$ \\
			\hline	
			\multicolumn{3}{c|}{${a}_2$} & $\ldots$ & $\ldots$ & $\ldots$ \\
			\multicolumn{3}{c|}{${a}_2$} & $\ldots$ & $\ldots$ & $\ldots$ \\
			& $\vdots$ &  & $\ldots$ & $\ldots$ & $\ldots$ \\
			\hline
			& $\vdots$ & & $\ldots$ & $\ldots$ & $\ldots$ \\

		\end{tabular}
		\vspace{2pt}
		\caption{The digit-table for a solution to $\Prob_N(p^\MC)$}
		
		\label{tab:conv-hnull-generic-ordering}
	\end{center}
\end{table}

The rows in each of the blocks in Table \ref{tab:conv-hnull-generic-ordering} are themselves arranged lexicographically, and so each block further splits into blocks depending on the values in the $l_0^*$ column. We may write block $i$ from Table \ref{tab:conv-hnull-generic-ordering} as in Table \ref{tab:conv-hnull-blocki-ordering}. The $\m_{ij}(M,\{l_0^*\})$ are not necessarily conforming digit-tables, but more on this in a moment.


\begin{table}[h]
	\begin{center}
		\begin{tabular}{ccc:c:ccc}
			
			$p^0$ & $p^1$ & $\cdots$ & $p^{l_0^*}$ & $p^{l_0^* +1}$ & $p^{l_0^* +2}$ \, $\cdots$ \, $p^{M-1}$\\
			\cline{1-6}
			\hspace{-.45in}\ldelim\{{14}{*}[$a_i$] & & & 0 & & &  \hspace{-.2in}\rdelim\}{3.5}{*}\\ 
			& & & \vdots & & \hspace{-.2in}$\m_{i0}(M,\{l_0^*\})$ & \\
			& & & 0 & & & \\
			\cdashline{1-6}
			& & & $1$ & & & \hspace{-.2in}\rdelim\}{3.5}{*}\\
		    & & & \vdots & & \hspace{-.2in} $\m_{i1}(M,\{l_0^*\})$ & \\
			& & & $1$ & & & \\
			\cdashline{1-6}
			& & & \vdots & & & \\
	        \cdashline{1-6}
	        & & & $p-1$ & & & \hspace{-.15in}\rdelim\}{3.5}{*}\\
		    & & & \vdots & & \hspace{-.2in}$\m_{i(p-1)}(M,\{l_0^*\})$ & \\
			& & & $p-1$ & & & 
		\end{tabular}
		\vspace{2pt}
		\caption{The structure of block $i$ from Table \ref{tab:conv-hnull-generic-ordering}}
		
		\label{tab:conv-hnull-blocki-ordering}
	\end{center}
\end{table}

Again, since a solution to $\Prob_N(p^\MC)$ is also a solution to $\Prob_N(p^{\mathcal{K}} )$ for any $\mathcal{K} \subseteq \MC$,
 applying the induction hypothesis to  $\MC_1 = \MC \setminus \{l_{k-1}\} $ as well as to $\{l_{k-1}\}$  the digit-table for any solution to $\Prob_N(p^\MC)$ is \emph{both} of the form
\begin{equation*}
\begin{pmatrix}
\vm(M, \MC_1') \\
\vm(M,\MC_1') \\
\vdots
\end{pmatrix} \quad \text{\emph{and}} \quad 
\begin{pmatrix}
\vm(M, \{l_0^\star\}) \\
\vm(M,\{l_0^\star\}) \\
\vdots
\end{pmatrix},
\end{equation*}
where $\MC_1' = M-1- \MC_1$ from the induction hypothesis. (Note that the pivot columns in the $\vm(M, \MC_1')$ are to the right of $l_0^\star$.) From this we can say the following:
\begin{enumerate}
 
	\item Since Table \ref{tab:conv-hnull-generic-ordering} needs to split into blocks $\vm(M, \{l_0^\star\})$,  each of the blocks in Table \ref{tab:conv-hnull-generic-ordering}  \emph{individually} must also split into blocks of conforming digit-tables $\vm(M, \{l_0^\star\})$. It is not possible for rows in different blocks of Table \ref{tab:conv-hnull-generic-ordering} to combine and form  a table $\vm(M, \{l_0^\star\})$ -- recall that all the initial columns preceding the first pivot column in any digit-table must contain the same digit.  As a consequence,  for the $i$'th block, above,  the $\m_{ij}(M,\{l_0^*\})$ in Table \ref{tab:conv-hnull-blocki-ordering}  are of the same size.

	\item Similarly, and in addition, since Table \ref{tab:conv-hnull-generic-ordering} also needs to split into blocks $\vm(M, \MC_1')$, each of the $\m_{ij}(M,\{l_0^*\})$ in Table \ref{tab:conv-hnull-blocki-ordering} must also split into blocks of conforming digit-tables $\vm(M, \MC_1')$. 
	
	From 1) since all the $\m_{ij}(M,\{l_0^*\})$ are equal in size, they split into the \emph{same} number of conforming sub-blocks. Let's say each $\m_{ij}(M,\{l_0^*\})$ splits into $s$ conforming sub-blocks $\vm_{ijk}(M,\MC_1')$, for $k = 0 $ to $s$. 
	Note that $\vm_{ijk}$ has $p^{|\MC|-1}$ rows.
\end{enumerate}

Now it is a simple matter of rearranging block $i$ in Table \ref{tab:conv-hnull-blocki-ordering} into $s$ sub-blocks, as in Table \ref{tab:conv-hnull-blocki-rearrangment}.

\begin{table}[h]
	\begin{center}
		\begin{tabular}{ccc|c|ccc}
			
			$p^0$ & $p^1$ & $\cdots$ & $p^{l_0^*}$ & $p^{l_0^* +1}$ & $p^{l_0^* +2}$ \, $\cdots$ \, $p^{M-1}$\\
			\cline{1-6}
			\hspace{-.45in}\ldelim\{{23.7}{*}[$a_i$] & & & 0 & &  &  \hspace{-.15in}\rdelim\}{2.5}{*}\\
			& & & $\vdots$ &  & \smash{\raisebox{.1in}{\hspace{-.5in}$\vm_{i00}(M,\MC')$}} 
			\\
			 \cdashline{1-6}
			  & & & 1 & & &    \hspace{-.15in}\rdelim\}{2.5}{*}\\
			 & & & $\vdots$ & & \smash{\raisebox{.1in}{\hspace{-.5in}$\vm_{i10}(M,\MC')$}}   \\
			 \cdashline{1-6}
			 & & & $\vdots$ & & & \\
	        \cdashline{1-6}
	        & & & $p-1$ & & &    \hspace{-.15in}\rdelim\}{2.5}{*}\\
			 & & & $\vdots$ & & \smash{\raisebox{.1in}{\hspace{-.5in}$\vm_{i(p-1)0}(M,\MC')$}}\\
			 \cline{1-6}
			 & & & 0 & & &    \hspace{-.15in}\rdelim\}{2.5}{*}\\
			 & & & $\vdots$ &  & \smash{\raisebox{.1in}{\hspace{-.5in}$\vm_{i01}(M,\MC')$}}\\
			 \cdashline{1-6}
			  & & & 1 & & &    \hspace{-.15in}\rdelim\}{2.5}{*}\\
			 & & & $\vdots$ & & \smash{\raisebox{.1in}{\hspace{-.5in}$\vm_{i11}(M,\MC')$}}\\
			 \cdashline{1-6}
			 & & & \vdots & & & \\
	        \cdashline{1-6}
	        & & & $p-1$ & & &    \hspace{-.15in}\rdelim\}{2.5}{*}\\
			 & & & $\vdots$ & & \smash{\raisebox{.1in}{\hspace{-.25in}$\vm_{i(p-1)1}(M,\MC')$}}\\
			 \cline{1-6}
			 
			 & & & \vdots & & & \\
	        \cdashline{1-6}

		\end{tabular}
		\vspace{2pt}

		\caption{The (rearranged) structure of block $i$ from Table \ref{tab:conv-hnull-generic-ordering}}
		
		\label{tab:conv-hnull-blocki-rearrangment}
	\end{center}
\end{table}
 
Note that the number of rows of each of the $s$ sub-blocks in Table \ref{tab:conv-hnull-blocki-rearrangment} is $p^{|\MC|}$. Thus each of the sub-blocks above is a conforming digit-table $\vm(M, \MC^*)$, and block $i$ in Table \ref{tab:conv-hnull-generic-ordering} is of the requisite form. Since this holds for any $i$ the entire digit-table is of the requisite form and the induction is complete.  This proves the necessity in Theorem~\ref{theorem:general-converse-hnull}. 

\subsubsection{Additional comments}
\label{sec:other-similar-results}
The study of vanishing sums of  roots of unity is a subject in itself, see e.g. \cite{conway1976trigonometric} and \cite{lam2000vanishing}, and underlies much of our work. Conceptually, Theorem \ref{theorem:general-converse-hnull} looks to be allied to results in the literature on minimal vanishing sums of roots, but to the best of our knowledge, the specific structure of an arbitrary solution to $\Prob_N(p^\L)$ and the style of proof presented here are new. The current approach also seems more amenable to generalizations when $N$ is not a prime power, very much a remaining challenge.

\bibliographystyle{IEEEtran}
\bibliography{OIS}
\end{document}